# A comparison of MIMO antenna efficiency measurements performed in Anechoic Chamber and Reverberation Chamber

*Tian-Hong Loh [1], and Wanquan Qi [2]*

[1] *National Physical Laboratory, Hampton Road, Teddington, Middlesex, TW11 0LW, UK*

[2] *Beijing Institute of Radio Metrology and Measurement, Yongdin Road, Haidian District, Beijing, China*

*Abstract* — Multiple-input-multiple-output (MIMO) antenna will play a key role in the development of fifth generation (5G) wireless mobile communication systems due to their performance-enhancement capability in multipath environment. Antenna radiation efficiency is an important parameter for MIMO antenna system. In this paper, we present a comparison of MIMO antenna efficiency measurements performed in Anechoic Chamber (AC) and Reverberation Chamber (RC) at the UK National Physical Laboratory. Two commercial available directional dual polarized full LTE band MIMO antennas were measured both in AC and RC between 1 GHz and 3 GHz.

## I. INTRODUCTION

Over the past two decades the market for modern wireless communication systems has grown rapidly in response to consumer demand. The use of multiple-input-multiple-output (MIMO) antennas, coupled with modulation formats, such as Orthogonal Frequency Division Multiple Access (OFDMA) can provide both increased channel capacity and protection against multi-path fading. This has been exploited and has encouraged, in recent years, extensive research activities in the application of MIMO technologies for these wireless systems, due to their rich scattering nature that provides improved spectral efficiencies and increased network capacity.

To date, MIMO technologies have been embedded in the fourth generation (4G) wireless communication systems such as 3GPP Long-Term-Evolution (LTE), WiMAX and Wi-Fi [1]. Furthermore, massive MIMO communication is an exciting area of fifth generation (5G) wireless research and it is envisaged that it becomes one of the major drivers for new radio access technologies in 5G [2].

In early literature on MIMO, the antenna elements were assumed isotropic and lossless [3]-[4]. However, in practice, they are often required to be compact for mobile communications where antenna characteristic such as antenna radiation pattern, antenna efficiency, correlation and mutual coupling cannot be omitted [5]-[6]. These are particularly important factors for the development of the 5G wireless communication systems as they promise to have seamless connectivity, which need to be further evolved beyond the current state of the art in order to accommodate a wide variety of issues and challenges. Therefore, system developers, manufacturers, and researchers need a good understanding of the real radiation characteristics from such devices in order to inform design and deployment choices.

Antenna radiation efficiency is one of the most important antenna parameter [7] as it would have significant effect on communications performance, reliability and efficiency of the system. It takes into account losses at the input terminals and within the structure of the antenna [8]. There are particular challenges in assessing the radiation efficiency of MIMO antennas as it has been demonstrated that both diversity gain and MIMO capacity depend upon the number of antennas, signal-to-noise ratio (SNR) and radiation efficiency on a complex way [9]. It should be noted that MIMO antenna radiation efficiency decreases by small array spacing where the reduction of efficiency reduces the channel capacity [10]. This will further pose additional difficulty and challenge for characterizing the antenna radiation efficiency of massive MIMO antenna.

The fundamental operating principles of anechoic chamber (AC) and reverberation chamber (RC) are different for antenna radiation efficiency measurements. AC is an ideal radio frequency free space environment, whereas RC is an over-moded reflecting environment providing statistically homogeneous and isotropic fields within its working volume. Antenna radiation efficiency measurements in AC are often affected by critical problems regarding measurement repeatability which is strongly dependent on the measurement configuration and the identification of the most sensitive parts of the antenna under test (AUT). Also an accurate reference gain antenna is required if antenna substitution method is employed for the AUT gain measurement. On the other hand, RC tests do not require accurate AUT positioning but require an accurate reference efficiency antenna.

There is limited number of literatures [11]-[12] presenting work relating to charactering MIMO antennas and comparing between AC and RC. Nevertheless, the main focus in [11]-[12] is on the characterization of the maximum-ratio-combining diversity gain and the ergodic MIMO capacity of MIMO antennas. The relevant AC diversity evaluations require measurement of embedded far-field functions and embedded radiation efficiencies at every antenna port. But considering the time consuming radiation pattern measurements in AC, the authors have assumed symmetric property of the MIMO antenna and only perform embedded far-field functions and embedded radiation efficiencies at one port. In this paper, we present a study into comparing the measured results obtained in AC and RC for two different directional dual polarized full LTE band MIMO antennas. The work aims to evaluate the antenna radiation efficiency at every antenna ports as well as the antenna radiation efficiency when the ports are combined using a combiner.

The paper is organized as follows: Section II presents the formula used for the radiation efficiency evaluation in AC and RC. Section III presents the details of measurement facilities and experimental setups for the radiation efficiency measurements of the MIMO antennas. Section IV shows a comparison between the radiation efficiency measurement results obtained from AC and RC. Finally, conclusions are drawn in Section V.

## II. THEORY

The following presents the formula used for the radiation efficiency evaluation in AC and RC, respectively.

### A. Radiation Efficiency using anechoic chamber

For AC, antenna radiation efficiency can be calculated using the antenna's gain and directivity. The directivity is a measure of the concentration of the radiation in a desired direction $(\theta_0, \emptyset_0)$ and the gain is the directivity including the losses up to the antenna output. These are defined as [13]:

$$\eta = G(\theta_0, \emptyset_0)/D(\theta_0, \emptyset_0). \qquad (1)$$

$$D(\theta_0, \emptyset_0) = F_{Nor}(\theta_0, \emptyset_0)/F_{Nor_{Av}}. \qquad (2)$$

$$F_{Nor_{Av}} = \frac{1}{4\pi} \iint_{4\pi} F_{Nor}(\theta, \emptyset)\, d\Omega. \qquad (3)$$

where $F_{Nor}(\theta_0, \emptyset_0)$ is the normalized radiation intensity for direction $(\theta_0, \emptyset_0)$, and $F_{Nor_{Av}}$ is the average value of normalized radiation intensity over $4\pi$ space.

### B. Radiation Efficiency using reverberation chamber

For RC, antenna radiation efficiency can be calculated using the following formula [14]:

$$\eta_{AUT} = F_{AUT} \cdot F_{REF} \cdot \eta_{REF} \qquad (4)$$

$$F_{AUT} = \frac{\langle |S_{21_{AUT}}|^2 \rangle}{\left(1 - |\langle S_{22_{AUT}} \rangle|^2\right)\left(1 - |\langle S_{22_{AUT}} \rangle|^2\right)} \qquad (5)$$

$$F_{REF} = \frac{\left(1 - |\langle S_{22_{REF}} \rangle|^2\right)\left(1 - |\langle S_{11_{REF}} \rangle|^2\right)}{\langle |S_{21_{REF}}|^2 \rangle} \qquad (6)$$

where $\eta$ represents the efficiency value. AUT and REF are for the antenna under test and the reference antenna, respectively.

## III. EXPERIMENTAL SETUP

All the radiation efficiency measurements were performed in the NPL AC and RC facilities. Two different commercially available directional dual polarized full LTE band two-port MIMO antennas were measured, namely, Laird PAS69278 and Poynting XPOL-A0002. An ETS-Lindgren 3117 double-ridged waveguide horn antenna was used as the Tx antenna in both AC and RC. The following presents the experimental setup for the radiation efficiency measurement in AC and RC, respectively. To assess the MIMO AUT as a single-port AUT a power combiner Mini-Circuits ZN2PD2-50-S+ was used. A four-port Rohde & Schwarz ZVB8 VNA was used with output power of 0 dBm.

### A. Anechoic Chamber

The anechoic chamber has a dimension of 7 m $\times$ 6.2 m $\times$ 6.2 m and has two low permittivity mounts, for the transmitting (Tx) and receiving (Rx) antennas (see Figs. 1(a) and 1(b), respectively). The roll-over-azimuth positioner system receiving end enables the three-dimensional (3D) radiation pattern of the Rx AUT located at the centre of rotation to be acquired over a spherical surface with a Tx probe antenna located at a fixed distance oriented transversely to the spherical surface with a particular polarization angle where vertical polarization (VP) and horizontal polarization (HP) are often been chosen.

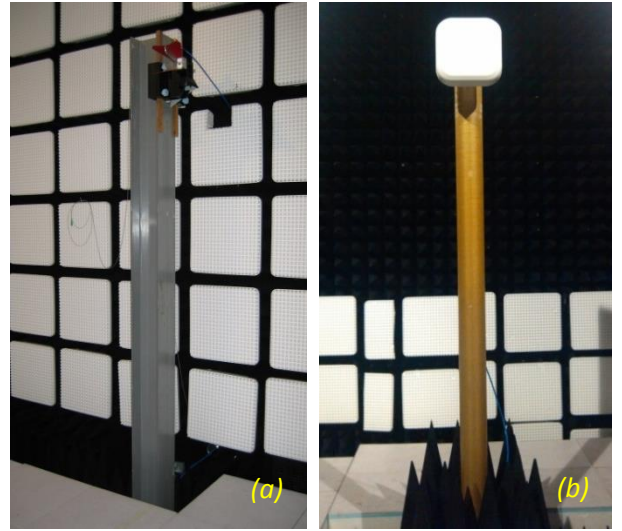

Fig. 1. NPL AC setup: (a) Transmitting tower with Tx ETS 3117 double-ridge horn antenna; (b) Receiving roll-over-azimuth positioner tower with the Laird PAS69278 MIMO AUT.

The distance between the Tx antenna and the Rx AUT was 2.8 m and the height of the antennas above the chamber ground was 3.058 m. Both the VP and HP radiation pattern measurements were performed with an angular resolution of 5° in the roll and azimuth spherical axes. An ETS-Lindgren 3117 double-ridged waveguide horn antenna was used as the reference gain antenna.

## B. Reverberation Chamber

The NPL RC has dimensions 6.5 m × 5.85 m ×3.5 m with spangle-galvanized mild stainless-steel walls and aluminium inner door skins (see Fig. 2). Three circular metallic plates are horizontally fixed to the vertical rotation paddle axis, which supports various rectangular aluminium tuner blades. An ETS-Lindgren 3117 double-ridged waveguide horn antenna was used as the reference efficiency antenna. Note that the stationary *S*-parameters was acquired for each of 1000 paddle steps (i.e. the paddle rotates a one-thousandth of a complete revolution).

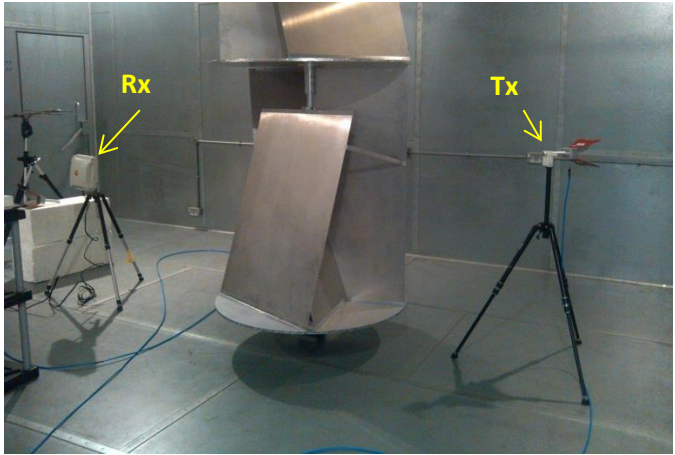

Fig. 2. . NPL RC setup for antenna radiation efficiency measurement.

## IV. Measurement Results

As shown in Figs. 3 and 4, the measurement results obtained in AC and RC are compared for the Laird PAS69278 and Poynting XPOL-A0002, respectively.

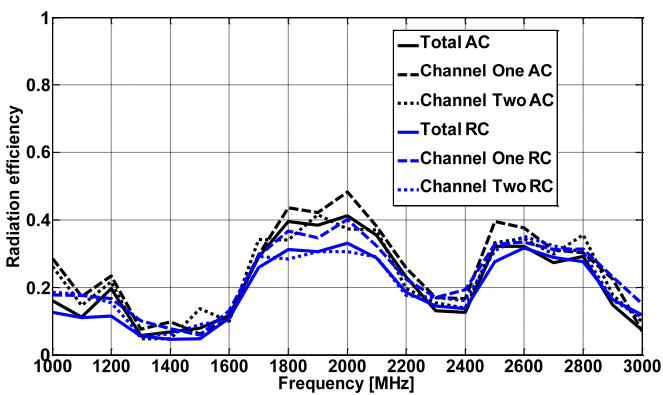

Fig. 3. Antenna radiation efficiency comparison between AC and RC for Poynting XPOL-A0002 AUT.

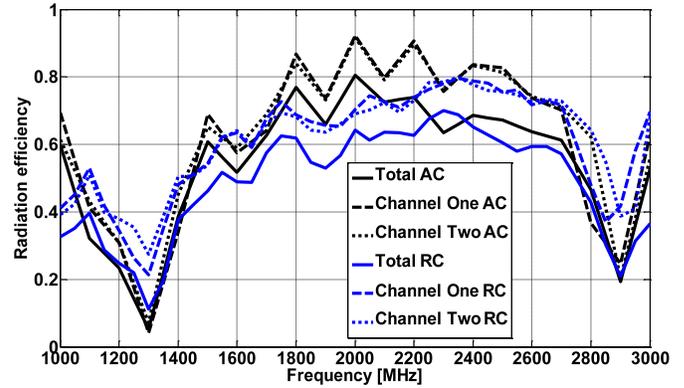

Fig. 4. Antenna radiation efficiency comparison between AC and RC for Laird PAS69278 AUT.

One notes that the legend text 'Total' means that the two-port are combined into one-port to connect to VNA (i.e. Port 1 at the VNA was connected to Tx and Port 2 at the VNA was connected to the combined port (using combiner) at Rx) whereas 'Channel One' and 'Channel Two' means that the two ports of the MIMO AUT are each connected to a port at VNA (i.e. Port 1 at the VNA was connected to Tx and Ports 2 and 3 at the VNA was connected to the two ports at Rx).

One observes from the above that the comparisons show reasonable agreements between AC and RC. Nevertheless, one notes that the antenna radiation efficiency is different at every antenna port as well as when the ports are combined.

## V. Conclusion

In this paper, we have presented a study into comparing the MIMO antenna efficiency measured results obtained in AC and RC for two different directional dual polarized two-port full LTE band MIMO antennas between 1 GHz and 3 GHz. We have demonstrated that the antenna radiation efficiency is different at every antenna ports as well as when the ports are combined. Considering that massive MIMO communication is envisaged to become one of the major drivers for new radio access technologies in 5G, one needs to further research on novel cost-effective measurement techniques to fully characterize its radiation efficiency.